\documentclass[reprint,twoside,superscriptaddress,aps,pra,longbibliography]{revtex4-2}
\usepackage[english]{babel}
\usepackage[utf8]{inputenc}
\usepackage[colorinlistoftodos, color=green!40, prependcaption]{todonotes}
\usepackage{siunitx}
\usepackage{svg}
\usepackage{graphicx}
\usepackage[caption=false]{subfig}
\usepackage[T1]{fontenc}
\newcommand\thefont{\expandafter\string\the\font}
\usepackage{amsmath}
\usepackage{mathtools}
\usepackage{mathrsfs}
\usepackage{mathptmx}
\usepackage{float}
\usepackage{tikz}
\usetikzlibrary{circuits.ee.IEC}
\usepackage{amssymb}

\begin{document}
%\title{Thermal response modeling of nanomechanical resonators for optimal photothermal signal estimation}
\title{Optimized Signal Estimation in Nanomechanical Photothermal Sensing via Thermal Response Modelling and Kalman Filtering}

\author{Hajrudin Bešić}\email{hajrudin.besic@tuwien.ac.at}% Your name
\affiliation{TU Wien, Institute of Sensor and Actuator Systems, Gusshausstrasse 27-29, 1040 Vienna, Austria}
\author{Andreas Deutschmann-Olek}
\affiliation{TU Wien, Automation and Control Institute, Gusshausstrasse 27-29, 1040 Vienna, Austria}
\author{Kenan Mešić}
\affiliation{TU Wien, Institute of Sensor and Actuator Systems, Gusshausstrasse 27-29, 1040 Vienna, Austria}
\author{Kostas Kanellopulos}
\affiliation{TU Wien, Institute of Sensor and Actuator Systems, Gusshausstrasse 27-29, 1040 Vienna, Austria}
\author{Silvan Schmid}
\affiliation{TU Wien, Institute of Sensor and Actuator Systems, Gusshausstrasse 27-29, 1040 Vienna, Austria}

\date{\today} % Leave empty to omit a date

\begin{abstract}
We present an advanced thermal response model for micro- and nanomechanical systems in photothermal sensing, designed to balance speed and precision. Our model considers the two time constants of the nanomechanical element and the supporting chip, triggered by photothermal heating, enabling precise photothermal input signal estimation through Kalman filtering. By integrating heat transfer and noise models, we apply an adaptive Kalman filter optimized for FPGA systems in real-time or offline. This method, used for photothermal infrared (IR) spectroscopy with nanomechanical resonators and a quantum cascade laser (QCL) in step-scan mode, enhances response speed beyond standard low-pass filters, reducing the effects of drift and random walk. Analytical calculations show the significant impact of the thermal expansion coefficients' ratio on the frequency response. The adaptive Kalman filter, informed by the QCL's input characteristics, accelerates the system's response, allowing rapid and precise IR spectrum generation. The use of the Levenberg-Marquardt algorithm and PSD analysis for system identification further refines our approach, promising fast and accurate nanomechanical photothermal sensing.
\end{abstract}

\keywords{photothermal spectroscopy, photothermal sensing, IR detection, nanomechanical resonators, NEMS, IR spectroscopy, Kalman Filter, Thermal Response}

\maketitle

\section{Introduction} \label{sec:intro}
%Nanomechanical resonators, which operate based on resonance frequency tracking, are employed for photothermal sensing with applications in infrared spectroscopy \cite{luhmann2023nanoelectromechanical} or as thermal infrared detector more general \cite{piller2022thermal}.

Nanomechanical resonators, pivotal for photothermal sensing through resonance frequency tracking, find extensive utility in fields such as infrared spectroscopy \cite{luhmann2023nanoelectromechanical,kurek2017nanomechanical} and broadband thermal infrared detection \cite{piller2022thermal,zhang2024high, Zhang2013, Duraffourg2018, Blaikie2019, Vicarelli2022, Li2023}. Demonstrated by these studies, nanomechanical photothermal sensing boasts low picogram and low picowatt sensitivity, respectively,  without the need for cryogenic cooling. Nanomechanical photothermal sensing is applicable to diverse samples such as nanoparticles \cite{yamada2013photothermal, kanellopulos2023nanomechanical}, pharmaceutical compounds \citep{kurek2017nanomechanical}, and polymer thin films \cite{casci2019thin}. However, nanomechanical resonators face two primary challenges: i) a response speed typically ranging from 5 ms to 100 ms, creating an intrinsic trade-off between speed and precision; and ii) a  step response dominated by two different transient responses, with the slower time constant often exceeding 100 seconds \cite{land2023sub}.

To address these challenges and enhance both the speed and precision of nanomechanical photothermal sensors, we develop a comprehensive thermal heat transfer and noise model, suited for optimal (Kalman) filtering. We test and apply the model and the Kalman-filter-based signal estimation by means of two experiments: one for model validation on a simple nanomechanical silicon-nitride string resonator, and another for conducting IR spectroscopy (NEMS-IR) on a more complex nanoelectromechanical drumhead resonator. 

In the two experiments, we employ pulsed laser light in the visible and mid-IR range of the electromagnetic spectrum, respectively.
%We employed a quantum cascade laser (QCL) to scan a range of the Mid-IR spectrum, thereby identifying the composition of polystyrene (PS) nanoparticles collected on the chip. The QCL was utilized in step-scan mode, where the laser's wavelength is adjusted, activated for a brief interval, then deactivated, and this process is repeated across the QCL's spectral range. 
The precise knowledge of the timing and shape of the input signal allows for adaptive Kalman filtering, significantly improving response speed compared to conventional low-pass filters. The designed Kalman filter estimates the input signal—specifically, the absorbed power—effectively bypassing the slow time constant characteristic of the measured signal. Additionally, by incorporating the system's noise profile, the filtering process is optimized for greater accuracy. System identification was achieved through a parameter fitting algorithm applied to a step response, adjusting initial parameters to align the modeled response with the actual measurement data.

We show that the presented model and Kalman filtering effectively minimizes the influence of slow, undesirable processes such as drift and random walk, paving the way for fast, precise nanomechanical photothermal sensing.

\section{Methods}

\subsection{Experimental Setups}
\subsubsection{String Resonator Setup}
\begin{figure}
    \centering
    \includegraphics[width=\columnwidth]{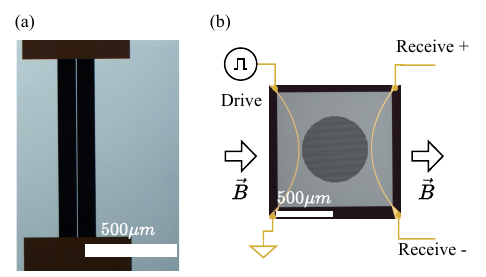}
    \caption{A microscope image of the nanomechanical resonators. (a) The silicon nitride string resonator. (b) The nanoelectromechanical drumhead resonator employed for infrared spectroscopy.}
    \label{fig:NEMS_resonator}
\end{figure}

To validate our model, we employ a nanomechanical silicon nitride string resonator, as shown in Figure \ref{fig:NEMS_resonator}a and detailed in prior work \cite{sadeghi2019influence}. The resonator has a thickness of \(56\,\mathrm{nm}\), a width of \(5\,\mathrm{\mu m}\), and a length of \(1\,\mathrm{mm}\). It operates under a tensile stress of approximately \(350\,\mathrm{MPa}\), ensuring optimal performance within a vacuum environment maintained at a pressure below \(10^{-5}\,\mathrm{mbar}\) to minimize air damping effects. The resonance frequency of the device is recorded at around \(170\,\mathrm{kHz}\), accompanied by a high quality factor (\(Q\)) of \(1.6\cdot10^6\) and a resonator time constant of \(2.5\,\mathrm{s}\). These parameters are integral to our experimental setup, providing a foundation for assessing the efficacy and accuracy of our proposed thermal heat transfer and noise model in a controlled environment.

\begin{figure*}
    \centering
    \includegraphics{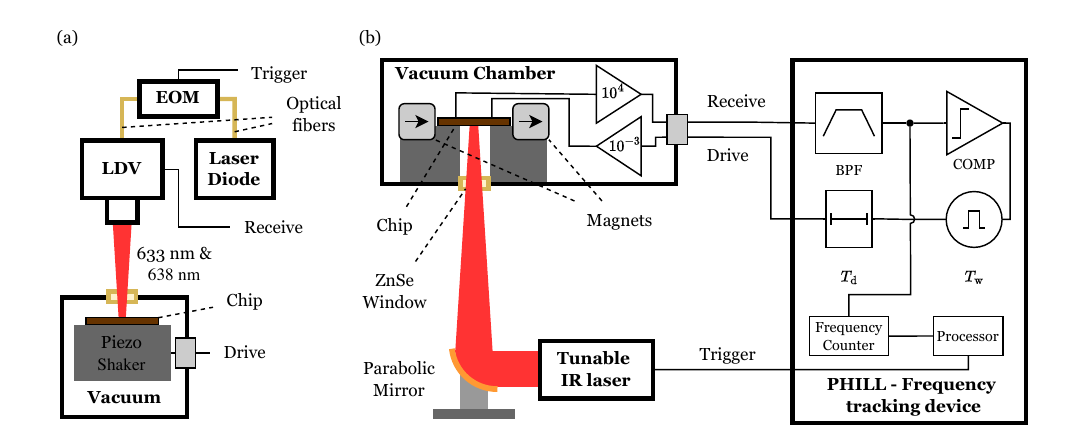}
    \caption{Schematic representation of the two measurement setups. Photothermal sensing with a) a nanomechanical string resonator in the visible and b) a nanoelectromechanical drumhead resonator to perform infrared spectroscopy NEMS-IR. Both setups are based on self-sustained frequency tracking with PHILL.}
    \label{fig:exp_setup}
\end{figure*}

\noindent The corresponding experimental setup is depicted in Figure \ref{fig:exp_setup}a. It utilizes a laser diode (LPS-635-FC from Thorlabs GmbH) with a wavelength of \(638\,\mathrm{nm}\) as the heating source. The laser diode is interconnected via an optical cable to an electro-optical modulator (EOM), which acts as a switch to create ideally shaped light pulses. The EOM is controlled by a "Trigger" signal, which is connected to our FPGA-based electronics (PHILL from Invisible-Light Labs GmbH). From the EOM, the laser is directed through an optical cable to a laser-Doppler vibrometer (LDV) (MSA-500 from Polytec GmbH). The LDV focuses the probing laser beam onto the nanomechanical string resonator and simultaneously measures its motion using an additional readout laser with a wavelength of \(633\,\mathrm{nm}\). The readout signal is then transmitted via the "Receive" line back to PHILL. The string resonator is actuated using a piezo shaker, which is driven by PHILL through the "Drive" line.

\subsubsection{NEMS-IR Setup} \label{sec:exp_setup}
%The two experimental setups used in this work, schematically shown in Figure~\ref{fig:exp_setup}, consist of three main parts: the vacuum chamber, laser source, and the frequency tracking device (PHILL from Invisible-Light Labs GmbH).

The NEMS-IR spectroscopy investigation employed a nanoelectromechanical resonator, which consists of a silicon nitride drumhead that is \SI{50}{\nano\meter} thick and measures \SI{1018}{\micro\meter} in length on each side, as detailed in prior work \cite{hajrudinSSO, hajrudin_FC} and depicted in Figure~\ref{fig:NEMS_resonator}b. For the purpose of electrical transduction, the setup includes two gold electrodes, each being \SI{5}{\micro\meter} in width, placed across the resonator. The system was positioned in a static magnetic field of about \SI{0.8}{\tesla}, produced by a Halbach array using neodymium magnets, with the electrode traces aligned perpendicularly to the field lines. This arrangement uses the Lorentz force, enabling one electrode to act as a driver by applying an alternating current, while the other electrode measured the resonator's movement via the induced voltage from magnetomotive forces. For signal detection enhancement, a differential pre-amplifier with a low noise profile and a gain of $10^4$ was utilized. The drumhead resonator operated in a vacuum environment, maintaining a pressure of $1.6\cdot10^{-5}\,\mathrm{mbar}$. It exhibited a resonance frequency of $f_\mathrm{r}=134$~kHz, quality factor of $13\,\mathrm{k}$ and resonator time constant of $31$~ms.

The experimental setup used to perform NEMS-IR is shown in \ref{fig:exp_setup}b. Inside the chamber, the chip is placed between the magnets of the Halbach array and above the ZnSe window that is located at the bottom of the vacuum chamber. The chip electrodes are contacted with spring-loaded contacts. The readout electrode pins are connected to the differential pins of the preamplifier. The output of the preamplifier is connected to the "Receive" line. One pin of the drive electrode is connected to ground. To limit the high current flowing through the chip and prevent damage, the other pin is connected over an attenuator, with gain $10^{-3}$, to the "Drive" line of the system.

To conduct NEMS-IR, we employ a tunable quantum cascade laser (QCL) as the laser source \cite{kurek2017nanomechanical, luhmann2023nanoelectromechanical}. Specifically, the MIRcat from Daylight Solutions laser is utilized, capable of generating up to \(500\,\mathrm{mW}\) of average laser power and scanning the laser wave-number within a range of \(1790\,\mathrm{cm}^{-1}\) to \(1122.5\,\mathrm{cm}^{-1}\). The laser's emission is focused onto the NEMS chip through a parabolic mirror and a ZnSe window. By sweeping the wave-number, the laser induces frequency detuning of varying magnitudes at different wave-numbers, thereby generating the IR spectrum.

The laser supports both continuous and step scan modes. In continuous mode, the laser remains activated while its wave-number is swept across the entire spectral range. Conversely, in step scan mode, the laser is set to a specific wave-number, activated, and after a predetermined hold time, deactivated and adjusted to a new wave-number. This sequence is reiterated until the full spectrum is covered. The step scan mode is preferred for its ability to generate a spectral sample by subtracting the "laser-off" frequency value from the "laser-on" value. Employing short sampling intervals allows for the suppression of slow processes such as thermal drift and random walk. The laser's state is synchronized with the PHILL frequency tracking device via the "Trigger" line, as illustrated in Figure~\ref{fig:exp_setup}b.

\subsubsection{Frequency Tracking}
The nanomechanical resonators are driven at their resonance frequency in a self-sustaining oscillator (SSO) resonance tracking scheme described in \cite{hajrudinSSO, Demir_tracking}. The feedback of the SSO is realized with the PHILL, as schematically depicted in Figure~\ref{fig:exp_setup}. The feedback of the closed-loop SSO scheme is realized with phase and amplitude controlling elements to satisfy the Barkhausen criterion. The feedback consists of a band-pass filter whose main task is mode selection of the measurement mode and suppression of unwanted modes. After the band-pass filter, a comparator with 50~mV hysteresis is placed to detect edges and act as a 0-degree phase detector. The output of the comparator is connected to the pulse-generating mechanism that generates timed pulses with a pulse width $T_\mathrm{w}$, delay relative to the detected phase of $T_\mathrm{d}$, and an amplitude $A$, which can be between 3~V and -3~V.  A frequency counter developed in \cite{hajrudin_FC} is used to acquire the frequency with a fixed sampling rate of around $20\,\mathrm{kHz}$.

\subsection{Thermal Response Model} \label{sec: thermal_resp}

The resonance frequency of a pre-stressed resonator is primarily influenced by the initial tension \cite{silvanbook2nd}. Variations in thermal expansion between the resonator and its supporting frame result in the resonator's strain being temperature-dependent, thereby inducing a temperature-dependent tensile stress. Consequently, the resonance frequency is also changed by temperature variations. Assuming a linear thermal expansion relationship for both the resonator and its frame, the temperature-induced strain in each of the two in-plane directions can be linearly approximated as \cite{silvanbook2nd}
\begin{equation}\label{eq:strain}
    \epsilon = \epsilon_0 - \alpha_\mathrm{r}\Delta T_\mathrm{r}+\alpha_\mathrm{f}\Delta T_\mathrm{f}
\end{equation}
where $\epsilon_0$ is the strain at the initial temperature $T_0$ without laser illumination, $\alpha_\mathrm{r/f}$ are the coefficients of thermal expansion of the resonator/frame and $\Delta T_\mathrm{r/f}=T_\mathrm{r/f}-T_0$ are the difference between the resonator/frame and the initial temperature. Solving (\ref{eq:strain}) for the temperature-dependent stress in each of the two in-plane directions, yields \cite{ventsel2002thin}
\begin{equation}
    \sigma(T) = \sigma_0 - \frac{E}{1-\nu}\left(\alpha_\mathrm{r} \Delta T_\mathrm{r} - \alpha_\mathrm{f} \Delta T_\mathrm{f}\right)
\end{equation}
where $\sigma_0$ is the stress at $T_0$, $E$ denotes Young's modulus, and $\nu$ is the Poisson's ratio. The fundamental mode eigenfrequency of a pre-stressed resonator can be calculated from the stress with the relation \cite{silvanbook2nd}
\begin{equation}
    \omega_\mathrm{0} = \frac{\pi}{L} \sqrt{\frac{\sigma_0}{\rho} -\frac{E(\alpha_\mathrm{r} \Delta T_\mathrm{r} - \alpha_\mathrm{f} \Delta T_\mathrm{f})}{\rho (1-\nu)}}
\end{equation}
where $L$ is the length and $\rho$ the mass density of the resonator. For small changes in temperature, the equation can be approximated by the first-order Taylor approximation
\begin{equation}
    \omega_\mathrm{0}(T) \approx \omega_\mathrm{0}(T_0)\left( 1-\frac{1}{2}     \frac{E(\alpha_\mathrm{r} \Delta T_\mathrm{r} - \alpha_\mathrm{f} \Delta T_\mathrm{f})}{\sigma_0 (1-\nu)} \right)
\end{equation}
In this simplified model, we ignore how temperature affects Young's modulus, thermal expansion coefficients, and mass density. However, in reality, a resonator's length and thermal expansion rates usually increase with temperature, while Young's modulus and mass density tend to decrease. However, these effects are negligible compared to the stress-change-induced frequency detuning. From the previous equation, the change in frequency caused by heating due to a change in temperature can be expressed as
\begin{equation}
    \Delta \omega_\mathrm{0} = - \omega_\mathrm{0} g \left( \alpha_\mathrm{r} \Delta T_\mathrm{r} - \alpha_\mathrm{f} \Delta T_\mathrm{f} \right) 
    \label{eq:freq_shift}
\end{equation}
where $g$ is a constant factor, which for a square drumhead is $g=E/[2 \sigma_0 (1-\nu)]$ and for a string with a uni-axial stress field reduces to $g=E/[2 \sigma_0]$. 

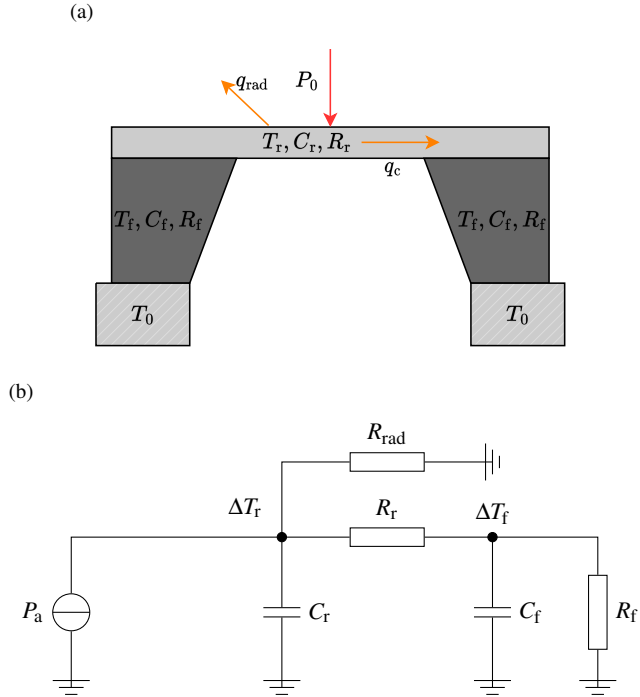
\begin{figure}
    \centering
    \captionsetup[subfigure]{position=top,singlelinecheck=off,justification=raggedright}

    \subfloat[]{\includegraphics[width=0.8\columnwidth]{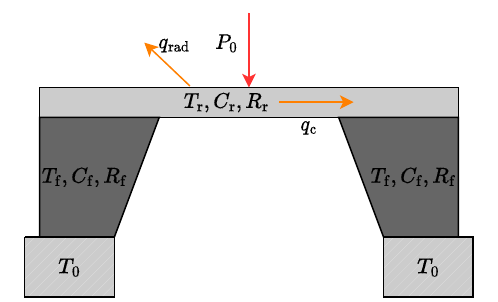}\label{fig:heat_flow}}
    
    \subfloat[]{
    \begin{tikzpicture}[circuit ee IEC]
        % Define parameters
        \def\vDist{2} % Vertical distance between components
        \def\hDist{2.8} % Horizontal distance between components
    
        % Ground
        \draw (0,0) node[ground,rotate=-90]{};
    
        % Draw source Rm and C_m
        \draw (0,0) to [current source = {info = $P_\mathrm{a}$}] (0,\vDist)
              to [] (\hDist,\vDist) 
              node[contact, label={[label distance=0.15cm]135:$\Delta T_\mathrm{r}$}]{}
              to [capacitor = {info = $C_\mathrm{r}$}] (\hDist,0)
              to (\hDist,0) node[ground,rotate=-90]{};
        %draw radiation source
        \draw (\hDist,  \vDist) -- (\hDist,1.5*\vDist) to [resistor = {info = $R_{\mathrm{rad}}$}] (2*\hDist,1.5 * \vDist) to (2*\hDist,1.5 * \vDist) node[ground,rotate=0]{};
        
        % Additional RC Circuit in parallel
        \draw (\hDist,\vDist) to [resistor = {info = $R_{\mathrm{r}}$}] (2*\hDist,\vDist)
              node[contact, label=above:$\Delta T_\mathrm{f}$]{}
              to [capacitor = {info = $C_{\mathrm{f}}$}] (2*\hDist,0)
              to (2*\hDist,0) node[ground,rotate=-90]{};
    
        % Add resistor to environment
        \draw (2*\hDist,\vDist) -- (2.5*\hDist,\vDist) 
              to [resistor = {info = $R_{\mathrm{f}}$}] (2.5*\hDist,0)
              to (2.5*\hDist,0) node[ground,rotate=-90]{};
    \end{tikzpicture}
    \label{fig:lumped-model}}
    
    \caption{(a) Schematic of the different parts of the resonator and frame contributing to the thermal circuit model. (b) Linearized thermal equivalent circuit model of the resonator and frame and its thermal connection to the environment. (The ground symbol represents the environment temperature $T_0$)}
    
\end{figure}

To utilize the static relation \eqref{eq:freq_shift}, the dynamics of $\Delta T_\mathrm{r}$ and $\Delta T_\mathrm{f}$ need to be described. Figure~\ref{fig:heat_flow} shows the schematic representation of the heat flow of the chip. Operating in vacuum, there are two dominant types of heat flow in the system: conduction and radiation \cite{piller2020thermal}. The resonator radiates heat on both sides and conducts heat over the frame into the environment. The heat flow can be approximated with a simple heat-transfer circuit diagram shown in Figure~\ref{fig:lumped-model}. The thermal equivalent circuit model consists of the heat source $P_\mathrm{a}$ and thermal resistances and capacitances of the resonator, frame, and chip holder. The absorbed power \( P_{\mathrm{a}} \) is calculated by multiplying the impinging power \( P_0 \) by the thermal absorption coefficient \( \beta \), as expressed by the equation 
\begin{equation}
P_{\mathrm{a}} = P_0 \beta.    
\end{equation}
The heat flow due to radiation can be modeled as a heat source described by the Stefan-Boltzmann law \cite{silvanbook2nd, piller2020thermal}
\begin{equation}
    q_\mathrm{rad} = 2 \eta A_\mathrm{r} \sigma_\mathrm{SB}[(\Delta T_\mathrm{r} +T_0)^4 - T_0^4].
\end{equation}
For small temperature changes $\Delta T_\mathrm{r}$, the radiation can be linearized around $T_0$ and modeled as a thermal resistance to the environment with a constant thermal resistance
\begin{equation}
    R_\mathrm{rad} = \frac{\Delta T_\mathrm{r}}{q_\mathrm{rad}} = 
    \frac{1}{8\eta A_\mathrm{r} \sigma_\mathrm{SB}T_0^3}
\end{equation}
where $\eta$ is the emissivity, $A_\mathrm{r}$ the surface of the resonator and $T_0$ the ambient temperature. From the thermal equivalent circuit model, a  linear state-space model can be extracted as
\begin{equation}
    \underbrace{
    \begin{bmatrix}
        \Delta \dot{T}_\mathrm{r} \\
        \Delta \dot{T}_\mathrm{f}
    \end{bmatrix}}_{\mathbf{\dot{x}_s}}
     = 
    \underbrace{
    \begin{bmatrix}
        -\frac{R_\mathrm{r}+R_\mathrm{rad}}{R_\mathrm{r}R_\mathrm{rad}C_\mathrm{r}} & \frac{1}{R_\mathrm{r}C_\mathrm{r}} \\
        \frac{1}{R_\mathrm{r}C_\mathrm{f}} & -\frac{R_\mathrm{r}+R_\mathrm{f}}{R_\mathrm{r}R_\mathrm{f}C_\mathrm{f}}  
    \end{bmatrix}}_{\mathbf{A_s}}
    \underbrace{
    \begin{bmatrix}
        \Delta {T_\mathrm{r}} \\
        \Delta {T_\mathrm{f}}  
    \end{bmatrix}}_{\mathbf{x_s}}
    +
    \underbrace{
    \begin{bmatrix}
        \frac{1}{C_\mathrm{r}} \\
        0
    \end{bmatrix}}_{\mathbf{B_s}}
    P_\mathrm{a} + \mathbf{w}
    \label{eq:state_eq}
\end{equation}
where $\mathbf{A_s}$ is the dynamic matrix, $\mathbf{B_s}$ the input matrix, $\mathbf{x_s}$ the state vector, $\mathbf{\dot{x}_s}$ its derivative and $\mathbf{w}$ the process noise in the system. From equation~\eqref{eq:freq_shift}, the corresponding output equation of the state-space model can be formed as follows
\begin{equation}
    \underbrace{\Delta \omega_\mathrm{0}}_y = 
    \underbrace{
    \begin{bmatrix}
        -\omega_\mathrm{0} g \alpha_\mathrm{r} &&
        \omega_\mathrm{0} g\alpha_\mathrm{f} 
    \end{bmatrix}}_{\mathbf{C_s}}
    \underbrace{\begin{bmatrix}
        \Delta {T_\mathrm{r}} \\
        \Delta {T_\mathrm{f}}  
    \end{bmatrix}}_\mathbf{x_s} + v
    \label{eq:out_eq}
\end{equation}
where $\mathbf{C_s}$ is the output matrix of the system, $y$ the measured quantity and $v$ the measurement noise. The state-space model from equation~\eqref{eq:state_eq} and \eqref{eq:out_eq}  can be discretized in time \cite{burns2001advanced} for a chosen sampling time $t_\mathrm{s}$ which yields
\begin{equation}
    \mathbf{x_s}(k+1)
     = 
    \underbrace{
        e^{\mathbf{A_s}t_\mathrm{s}}
        }_{\mathbf{F_s}}
    \mathbf{x_s}(k)
    +
    \underbrace{
      \int_0^{t_\mathrm{s}} e^{\mathbf{A_s} \tau} \mathbf{B_s}d\tau
      }_{\mathbf{G_s}}
    P_\mathrm{a}(k)   + \mathbf{w}(k),
    \notag
\end{equation}

\begin{equation}
    y(k) = 
    \underbrace{
        \mathbf{C_s}
    }_{\mathbf{H_s}}
    \mathbf{x_s}(k) + v(k)
    \label{eq:state_eq_d}
\end{equation}
where $\mathbf{F_s}$ is the discrete dynamic matrix, $\mathbf{G_s}$ the discrete input matrix and $\mathbf{H_s}$ is the discrete output matrix. The discrete state-space model \eqref{eq:state_eq_d} is particularly well suited for signal processing on digital systems.

\section{Noise Modeling and Kalman Filtering}

Noise modeling is a crucial aspect of Kalman filtering, an optimal estimation technique widely used in control systems, navigation, and signal processing \cite{kalman1960new, welch1995introduction}. The primary goal of Kalman filtering is to estimate the state of a dynamic system from a series of noisy measurements. To achieve this, the Kalman filter relies on accurate models of both the system dynamics and the noise affecting the system. Specifically, the noise is categorized into two types: measurement noise and process noise, described by $\mathbf{w}$ and $v$ in equations~\eqref{eq:state_eq} and \eqref{eq:out_eq}. Measurement noise represents the errors in the observations \cite{brown1983introduction, gelb1974applied}, while process noise accounts for the uncertainties in the system model itself \cite{ribeiro2004kalman}. The measurement variance (\(R\)) quantifies the expected errors in the measurements with the expected value $R(t-\tau) = \mathbb{E}[v(t)v(\tau)]$, and the process covariance matrix (\(\mathbf{Q}\)) characterizes the uncertainties in the system dynamics with the expected value $\mathbf{Q}(t-\tau) = \mathbb{E}[\mathbf{w}(t)\mathbf{w}^\mathrm{T}(\tau)]$. These covariance matrices are essential for the Kalman filter to weigh the reliability of the predictions versus the new measurements correctly. By accurately modeling the noise, the Kalman filter can optimally combine information from the model and the measurements, resulting in precise and reliable state estimates. Understanding and implementing noise models is therefore fundamental to the effective application of Kalman filtering \cite{anderson1979optimal, simon2006optimal}. In the following, we will describe the measurement and process noise of the system, disregarding slow processes such as drift and random walk. These slow noise processes are irrelevant because sampling occurs on much smaller timescales.

\subsubsection{Measurement Noise}
\begin{figure}
    \centering
    \includegraphics{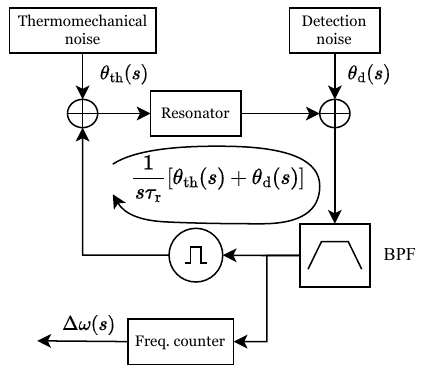}
    \caption{Noise contributions in the SSO frequency tracking scheme, represented in phase space}
    \label{fig:noise-block}
\end{figure}

The measurement noise refers to the disturbances generated during the measurement of a specific physical quantity, in this instance, the temperature of the resonator.  
For a measurement performed in an SSO configuration two main noise contributors can be identified as shown in Figure~\ref{fig:noise-block}: thermomechanical noise and detection noise. The thermomechanical noise is fundamental to nanomechanical resonators and can be modeled as white force noise acting on the resonator. This causes an amplitude noise to the position of the resonator at resonance with the power spectral density (PSD) \cite{Demir_tracking,silvanbook2nd, hajrudinSSO} of 
\begin{equation}
    S_\mathrm{thm} = \frac{4 k_\mathrm{B}T_\mathrm{r} Q}{m \omega_0^3}
\end{equation}
where $k_\mathrm{B}$ is Boltzmann constant, $T_\mathrm{r}$ the temperature, $m$ the effective mass,  $\omega_0$ the eigenfrequency, and $Q$ is the quality factor of the resonator.

Detection noise is produced by transducing the mechanical motion of the resonator into an electrical signal and is caused mostly by transduction and electronic noise. Being typically white Gaussian noise, detection noise PSD can be described with respect to the thermomechanical noise
\begin{equation}
    S_\mathrm{d} = \mathcal{K}^2 S_\mathrm{thm},
\end{equation}
where $\mathcal{K}$ is a dimensionless factor that corresponds to the ratio between the detection noise background and the height of the thermomechanical noise peak. The amplitude domain PSDs can be transformed into phase space by \cite{Demir_tracking, rubiola2008phase}
\begin{equation}\label{eq:phasenoise}
\begin{split}
    S_{\theta_\mathrm{thm}} &= \frac{2}{A_\mathrm{r}^2} S_\mathrm{thm}\\
    S_{\theta_\mathrm{d}} &= \frac{2}{A_\mathrm{r}^2} S_\mathrm{d} = \mathcal{K}^2 S_{\theta_\mathrm{th}}
\end{split}
\end{equation}
with the vibrational amplitude at resonance $A_\mathrm{r}$.

The measurement noise sources contribute to the measuring noise in two different ways. The first is caused by phase detection noise passing through the band-pass filter (BPF), then through the low-pass filter of the frequency counter, and subsequently being transformed into frequency through differentiation. The second contributor is the closed-loop noise caused by the thermo-mechanical and detection noise combined. This type of noise is described by the Leeson effect \cite{rubiola2008phase,Demir_tracking,hajrudinSSO}, which asserts that within an SSO tracking scheme, phase noise is integrated and is proportional to the reciprocal of the resonator's time constant $\tau_\mathrm{r}$, provided that it is significantly larger than the inverse of the band-pass filter's bandwidth. This integrated noise, after passing through the band-pass filter and frequency counter filter, is then differentiated and transformed into frequency. It can be described by white Gaussian thermo-mechanical noise passing through the transfer function $G_\mathrm{v}$ yielding the power spectral density
\begin{equation}
    S_\mathrm{v}(s) = |G_\mathrm{v}(s)|^2 S_\mathrm{\theta_\mathrm{thm}}(s),
    \label{eq:S_v}
\end{equation}
with
\begin{equation}
    G_\mathrm{v}(s)=\frac{s\mathcal{K} + \frac{\sqrt{\mathcal{K}^2+1}}{\tau_\mathrm{r}}}{(1+s\cdot \tau_\mathrm{BP})(1+s \cdot \tau_\mathrm{FC})}
\end{equation}
where $\tau_\mathrm{BP}$ represents the time constant of the band-pass filter, and $\tau_\mathrm{FC}$ is the time constant of the frequency counter. Considering that the measurement noise follows a Gaussian distribution (even though it has not a constant PSD), a straightforward approach would be to empirically measure the noise variance and use
\begin{equation}
    R=\sigma_\mathrm{v}^2.
\end{equation}

Literature \cite{brown1983introduction, popescu1998kalman} suggests that a better approach for treating colored noise is to build a noise model based on its transfer function and augment it to the system in \eqref{eq:state_eq}. The noise state space can be described in controllable canonical form as
\begin{subequations}\label{eq:noise_eq_d}
\begin{equation}
    \mathbf{\dot{x}_{v}}(t)
     = 
     \underbrace{
    \begin{bmatrix}
        0 & 1 \\
        -\frac{1}{\tau_\mathrm{BP}\tau_\mathrm{FC}} & - \left( \frac{1}{\tau_\mathrm{BP}} + \frac{1}{\tau_\mathrm{FC}} \right) 
    \end{bmatrix}}_\mathbf{A_{v}}
    \mathbf{x_{v}}(t)
    +
    \underbrace{
    \begin{bmatrix}
        0 \\
        1
    \end{bmatrix}}_\mathbf{B_{v}} u_\mathrm{v}(t),
\end{equation}
\begin{equation}
    v(t) =
    \underbrace{
    \begin{bmatrix}
        \frac{\sqrt{\mathcal{K}^2+1}}{\tau_\mathrm{BP}\tau_\mathrm{FC}} &&
        \frac{\mathcal{K}}{\tau_\mathrm{BP}\tau_\mathrm{FC}}
    \end{bmatrix}}_\mathbf{C_{v}}
    \mathbf{x_{v}}(t),
\end{equation}
\end{subequations}
where $u_\mathrm{v}$ is the white measurement noise sequence with its auto-correlation  $\mathbb{E}[u_\mathrm{v}(t)u_\mathrm{v}(\tau)] = S_{\theta_\mathrm{thm}}\delta(t-\tau)$, where $\delta(\cdot)$ is the Dirac delta function. The noise model can be discretized in time as shown in equation~\eqref{eq:state_eq_d} yielding the discrete-time matrices $\mathbf{F_{v}}$, $\mathbf{G_{v}}$ and $\mathbf{H_{v}}$. 

%The discrete process noise covariance matrix for the measurement noise state space can be written as \cite{kulikov2013accurate}
%\begin{equation}
    %\mathbf{Q_{v}} \approx 
    %\mathbf{G_{v}} S_\mathrm{v} t_\mathrm{s} \mathbf{G^T_{v}} 
%.
%\end{equation}

\subsubsection{Process Noise}
The process noise represents the disturbances acting on the modeled states. In this model, it is epitomized by the temperature fluctuations exerted on the resonator. Temperature fluctuations are a fundamental phenomenon affecting all structures, primarily arising from the statistical and quantum aspects of heat exchange \cite{vig1999noise}. This process involves the absorption and emission of photons at the resonator surfaces, alongside thermal conduction to and from the surrounding environment via phonons. The resultant noise from this heat exchange manifests as shot noise, which exhibits white frequency noise behavior filtered by the resonator thermal response.

In the context the system described in \eqref{eq:state_eq}, two temperature states are modeled: the resonator and the frame. The frame, due to its large size and very high thermal capacity, is assumed to have a static temperature with negligible fluctuations. Conversely, the resonator experiences significant temperature fluctuations, described by the spectral density \cite{vig1999noise}:
\begin{equation}
    S_\mathrm{\Delta T_\mathrm{r}}(s) = \bar{S}_\mathrm{\Delta T_\mathrm{r}} |G_\mathrm{th}(s)|^2 = 4 k_\mathrm{B} T_0^2 R_\mathrm{th}|G_\mathrm{th}(s)|^2,   
\end{equation}
where \( R_\mathrm{th} = R_\mathrm{r} \parallel R_\mathrm{rad} \) is the equivalent thermal resistance, and
\begin{equation}
    G_\mathrm{th}(s) = \frac{1}{1+s \cdot R_\mathrm{th} C_\mathrm{r}}
    \label{eq:thermal_tf}    
\end{equation}
is a first-order low-pass filter determined by the resonator's thermal properties. This behavior, when expressed in the controllable canonical state space form, can be articulated as follows
\begin{subequations}\label{eq:noise_eq_pn}
\begin{equation}
    \dot{x}_{w}(t)
     = 
    \underbrace{
    \begin{bmatrix}
        \frac{1}{\tau_\mathrm{th}}
    \end{bmatrix}}_{{A_\mathrm{w}}}
    {x_\mathrm{w}}(t)
    +
    \underbrace{
    \begin{bmatrix}
        1
    \end{bmatrix}}_{{B_\mathrm{w}}}
    u_\mathrm{w}(t),
\end{equation}
\begin{equation}
    \mathbf{w}(t) = 
    \underbrace{
    \begin{bmatrix}
        \frac{1}{\tau_\mathrm{th}} \\ 0
    \end{bmatrix}}_{\mathbf{C_{w}}}
    {x_\mathrm{w}}(t).
\end{equation}
\end{subequations}
where $u_\mathrm{w}$ is the white measurement noise sequence with its auto-correlation $E[u_\mathrm{w}(t)u_\mathrm{w}(\tau)] = \bar{S}_\mathrm{\Delta T_\mathrm{r}}\delta(t-\tau)$. The state space can again be discretized as shown in equation~\eqref{eq:state_eq_d}  yielding the discrete state space model ${F_\mathrm{w}}$, ${G_\mathrm{w}}$ and $\mathbf{H_{w}}$. %The discrete process noise covariance matrix is expressed as \cite{kulikov2013accurate}
%\begin{equation}
%    \mathbf{Q_{w}} \approx 
%    \mathbf{G_{w}} S_\mathrm{\Delta T_\mathrm{r}} t_\mathrm{s} \mathbf{G^T_{w}} 
%.
%\end{equation}
Combining the system dynamics \eqref{eq:state_eq_d} with the discretized colored noise models \eqref{eq:noise_eq_d} and \eqref{eq:noise_eq_pn} yields
\begin{subequations}\label{eq:noise_eq_sys}
\begin{align}
    \mathbf{x_s}(k+1) &= \mathbf{F_s} \mathbf{x_s}(k) + \mathbf{G_s} P_\mathrm{a}(k) + \mathbf{H_{w}} {x_\mathrm{w}}(k)\\
    {x_\mathrm{w}}(k+1) &= {F_\mathrm{w}} {x_\mathrm{w}}(k) + {G_\mathrm{w}}u_\mathrm{w}(k)\\
    \mathbf{x_{v}}(k+1) &= \mathbf{F_{v}} \mathbf{x_{v}}(k) + \mathbf{G_{v}}u_\mathrm{v}(k)\\
    y(k) &= \mathbf{H_s} \mathbf{x_s}(k) + \mathbf{H_{v}} \mathbf{x_{v}}(k),
\end{align}
\end{subequations}
resulting in a more comprehensive model that accounts for both measurement and process noise profiles.
% resulting in an augmented state model
% \begin{equation}
%     \mathbf{x} = \begin{bmatrix}
%         \mathbf{x_s} \\
%         \mathbf{x_v} \\
%         \mathbf{x_w}
%     \end{bmatrix},
%     \mathbf{F} = 
%     \begin{bmatrix}
%         \mathbf{F_s} & 0 & \mathbf{H^T_{w}} \\
%         0 & \mathbf{F_{v}} & 0\\
%         0 & 0 & \mathbf{F_{w}}
%     \end{bmatrix}, 
%     \mathbf{G} = 
%     \begin{bmatrix}
%         \mathbf{G_s} & 0 & 0\\
%         0 & \mathbf{G_{v}} & 0\\
%         0 & 0 & \mathbf{G_{w}}
%     \end{bmatrix}, 
%     \notag
% \end{equation}
% \begin{equation}
%     \mathbf{H} = 
%     \begin{bmatrix}
%         \mathbf{H_s} & \mathbf{H_{v}} & 0
%     \end{bmatrix},
%     \mathbf{Q}= 
%     \begin{bmatrix}
%         0 & 0 & 0\\
%         0 & \mathbf{Q_{v}} & 0\\
%         0 & 0 & \mathbf{Q_{w}}
%     \end{bmatrix}.
% \end{equation}
%The measurement noise has been transferred from the measurement equation to the state equation. Consequently, this shift moves the noise from the measurement covariance matrix to the process noise covariance matrix, resulting in $\mathbf{R}=[0]$.

\subsection{Adaptive Kalman Filter} \label{sec: kalman_filter}
The fundamental speed versus accuracy limit of nanomechanical resonators is intrinsic. The accuracy can be improved by noise filtering, at the cost of speed reduction \cite{hajrudinSSO,Demir_tracking}. However, this limit can be broken using additional information about the system. In the case of NEMS-IR using QCL step scan mode, it is known that the power input has a step shape with unknown height and known position in time (over the trigger in Figure~\ref{fig:exp_setup}). It can be modeled as an unknown but constant power with a small disturbance term $w_\mathrm{P_\mathrm{a}}$ to introduce a form of uncertainty
\begin{equation}
    P_\mathrm{a}(k+1) = P_\mathrm{a}(k) +w_\mathrm{P_\mathrm{a}}(k).
    \label{eq:const_pow_eq}
\end{equation}
Through the thermal equivalent circuit model and equations~\eqref{eq:state_eq} and \eqref{eq:out_eq}, the dynamics of the system are also well known. For problems of this type, an adaptive Kalman filter (AKF) can be used \cite{demir2021adaptive,verhaegen2007filtering}. Given the knowledge of the model, inputs, and stochastic properties of the system, it is capable of estimating the system in a statistically optimal manner \cite{verhaegen2007filtering}. Despite the knowledge of the input timing and shape the height of the absorbed laser power is unknown. Therefore, the model has to be extended and the Kalman filter needs to estimate in addition to the states the value of $P_\mathrm{a}$. This can be achieved by augmenting the model \cite{verhaegen2007filtering}, by adding \eqref{eq:const_pow_eq} to the system \eqref{eq:noise_eq_sys} yielding
\begin{subequations}\label{eq:state_eq_d_aug}
\begin{equation}
    \begin{aligned}
    \underbrace{
    \begin{bmatrix}
    \mathbf{x_s}({k+1}) \\
    \mathbf{x_v}(k+1) \\
    x_\mathrm{w}(k+1) \\
    P_\mathrm{a}({k+1})
    \end{bmatrix}}_{\mathbf{x}(k+1)}
    &=
    \underbrace{
    \begin{bmatrix}
    \mathbf{F_s} & 0 & \mathbf{H_w} & \mathbf{G_s} \\
    0 & \mathbf{F_v} & 0 & 0 \\
    0 & 0 & \mathbf{F_w} & 0\\
    0 & 0 & 0 & 1
    \end{bmatrix}}_\mathbf{F}
    \underbrace{
    \begin{bmatrix}
    \mathbf{x_s}({k}) \\
    \mathbf{x_v}(k) \\
    x_\mathrm{w}(k) \\
    P_\mathrm{a}({k})
    \end{bmatrix}}_{\mathbf{x}(k)} \\
    &\quad+
    \underbrace{
    \begin{bmatrix}
        0 & 0 & 0 \\
        \mathbf{G_v} & 0 & 0\\
        0 & G_\mathrm{w} & 0 \\
        0 & 0 & 1
    \end{bmatrix}}_\mathbf{G}
    \underbrace{
    \begin{bmatrix}
    u_\mathrm{v}(k) \\
    u_\mathrm{w}(k) \\
    w_\mathrm{P_\mathrm{a}}(k)
    \end{bmatrix}}_\mathbf{\bar{w}},
    \end{aligned}
\end{equation}
\begin{equation}
    y(k) = 
    \underbrace{
    \begin{bmatrix}
        \mathbf{H_s} & \mathbf{H_v} & 0 & 0
    \end{bmatrix}}_\mathbf{H}
    \mathbf{x}(k).
\end{equation}
\end{subequations}
%where \( w_\mathrm{P_\mathrm{a}}(k) \) represents the estimation error of the unknown power state. 
Notice that the measurement noise has been transferred from the output equation to the state equation. As a result, the system does not have measurement noise covariance matrix and all noise prcesses are described by the process noise covariance matrix. This matrix, described in the discrete time domain, is given by

\begin{equation}
    \mathbf{Q} = \mathbb{E}[\mathbf{\bar{w}}\mathbf{\bar{w}}^\mathrm{T}] = 
    \begin{bmatrix}
         S_\mathrm{v} t_\mathrm{s} & 0 & 0 \\
        0 & \bar{S}_\mathrm{\Delta T_\mathrm{r}} t_\mathrm{s} & 0 \\
        0 & 0 & \sigma^2_{P_\mathrm{a}}
    \end{bmatrix}.
\end{equation}
where $\sigma^2_\mathrm{P_\mathrm{a}}$ is the process covariance matrix of the estimated absorbed power. The variance $\sigma^2_\mathrm{P_\mathrm{a}}$ is a tuning factor and has to be determined empirically. The Kalman filter algorithm is described in \cite{demir2021adaptive} and consists of the following steps
\begin{subequations}
\begin{itemize}
    \item {\bf Predict}
    \begin{align}
        \hat{\mathbf{x}}(k|k-1) &= \mathbf{F} \hat{\mathbf{x}}(k-1|k-1), \\
        \mathbf{P}(k|k-1) &= \mathbf{F} \mathbf{P}(k-1|k-1) \mathbf{F}^T + \mathbf{G}\mathbf{Q}\mathbf{G}^\mathrm{T}, 
    \end{align}
        
    \item {\bf Observe}
    \begin{align}
        \tilde{y}(k) &= y(k) - \mathbf{H} \hat{\mathbf{x}}(k|k-1), \\
        \mathbf{V}(k) &= \mathbf{H} \mathbf{P}(k|k-1) \mathbf{H}^T + \mathbf{R},
    \end{align}

    \item {\bf Kalman gain}
    \begin{align}
        \mathbf{K}(k) &= \mathbf{P}(k|k-1) \mathbf{H}^T \mathbf{V}^{-1},
    \end{align}

    \item {\bf Estimate}
    \begin{align}
        \hat{\mathbf{x}}(k|k) &= \hat{\mathbf{x}}(k|k-1) + \mathbf{K}(k) \tilde{y}(k), 
    \end{align}

    \item {\bf Estimate covariance}
    \begin{align}
        \mathbf{P}(k|k) &= (\mathbf{I} - \mathbf{K}(k) \mathbf{H}) \mathbf{P}(k|k-1), 
    \end{align}
\end{itemize}
\end{subequations}
where $\hat{\mathbf{x}}$ is the estimated state, $\tilde{y}$ the difference between the measured and predicted state also called the innovation or residual, $\mathbf{P}$ the covariance matrix of the state estimate and $\mathbf{K}$ the Kalman gain. The notation $(n|m)$ means that at the corresponding quantity is estimated at a time $n$ from the time $m$ that is located in the past, i.e. $m \leq n$. The Kalman filter is a recursive algorithm used to estimate the state of a dynamic system from a series of noisy measurements. It combines predictions from a model with new measurements to produce an estimate that minimizes the mean of the squared error. The covariance matrix \(\mathbf{P}\) represents the uncertainty in the state estimate. It is updated at each step to reflect the confidence in the estimate.

%To make the Kalman algorithm adaptive the trigger information is used. When the trigger signal goes from low to high of from high to low, it means that the laser is turned on or off. This means that the Kalman filter should adapt more quickly because an event happened. To accomplish this the covariance matrix $\mathbf{P}$ is manually increased, which will increase the Kalman gain and the filter will adapt to the new incoming measurements more quickly but also filter less measurement noise. After a while the filter will converge back to its original filtering strength. 

When ever the laser is switched on or off, estimates of $P_\mathrm{a}$ from previous data are clearly becoming invalid. This can easily be included in the adaptive Kalman filter by manualy increasing the covariance of the estimate $\hat{P}_\mathrm{a}$ (or even the whole covariance matrix $\mathbf{P}$). This way, the AKF adapts quickly to new measurement information at expense of more noisy estimates in the beginning. After a while the filter will converge back to its original filtering strength. The estimate of the impinging power $\hat{P}_0$ is obtained, by simply dividing the absorbed power estimate $\hat{P}_\mathrm{a}$ with the absorption coeficient of the structure $\beta$.

\subsection{System Identification} \label{sec: sys_ident}
Mathematical models are often derived from idealized scenarios and need to be calibrated to reproduce experimental data. In real world applications, factors like chip fabrication, laser alignment, or spot size of the laser illuminating the membrane can vary widely. Provided the system is linear, a step response system identification can be a powerful tool to properly characterize the different parameters of a dynamical system. The method of least squares is  widely regarded as a simple and effective technique for extracting information from a dataset \cite{wolberg2006data}. The method is optimal in the sense that the parameters determined by the least squares analysis are normally distributed about the true parameters with the least possible standard deviations.
 
The method of least squares is a parameter estimation method in regression analysis based on minimizing the sum of the squares of the residuals (a residual being the difference between an observed value and the fitted value provided by a model) made in the results of each individual equation. However, problems with this method can arise if the numerical algorithm gets stuck in a local minimum of the error function, resulting in suboptimal parameters. To prevent this from happening, in this case, initial parameters are roughly calculated analytically. They ensure that the numerical algorithm doesn't start too far from the global minimum, thus increasing its chances of finding the optimal solution. In this work, the {\it Levenberg–Marquardt } algorithm was used due to its robustness.

\section{Results and Discussion} \label{sec:results}

\begin{figure}
    \centering
    \includegraphics[width=0.4\textwidth]{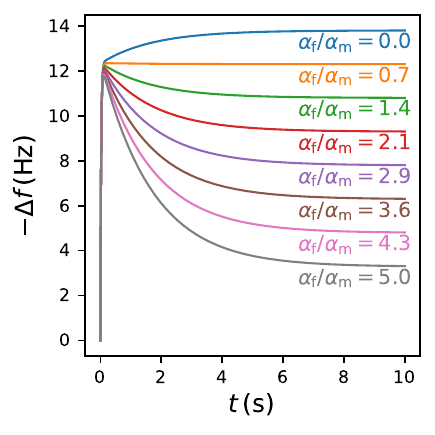}
    \caption{Step-responses of the system for different ratios of $\alpha_\mathrm{f}$ and $\alpha_\mathrm{m}$.}
    \label{fig:alpha-sweep}
\end{figure}
Thermal systems that operate in a linear regime usually do not exhibit overshoots. It is mentioned in section~\ref{sec: thermal_resp} that it is possible that the deformation of the frame can cause this effect due to its thermal expansion coefficient $\alpha_\mathrm{f}$. Figure~\ref{fig:alpha-sweep} shows step responses generated for different ratios of the thermal coefficient of the membrane and frame. It can be seen that the slow frame-induced frequency shift does not have to always oppose the fast one. Usually, if $\alpha_\mathrm{f} > c\alpha_\mathrm{m}$ the frame expansion will oppose the fast frequency shift, and for $\alpha_\mathrm{f} < c\alpha_\mathrm{m}$ it will act in the same direction. In this case, $c \approx 0.7$ represents the equilibrium point where frame stretching and heating effects are balanced, resulting in a minimal impact on the response. It is important to note that, in theory, ratios exist where the effect of the frame can completely diminish initial membrane-induced frequency shift and at steady state $\Delta f = 0$. That means a mathematical configuration exists where the resonator acts as a high-pass filter. The other extreme theoretical case is when $\alpha_\mathrm{f} = 0$. In this case the frame is not stretching and only heating. Then the steady-state value can be a lot larger than the initial jump from the fast process. This could be used as an advantage only if the drift of the sensor and other slow processes could be managed.

\subsection{String Resonator Setup}
\begin{figure*}
    \centering
    \captionsetup[subfigure]{position=top,singlelinecheck=off,justification=raggedright}

        \subfloat[]{\includegraphics{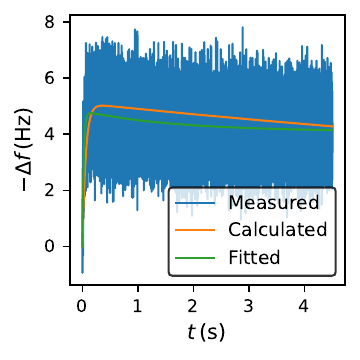}\label{fig:string_fit}}
        \subfloat[]{\includegraphics{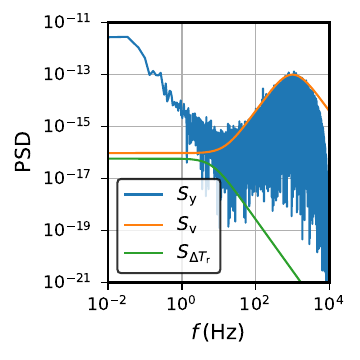}\label{fig:string_PSD}}
        \subfloat[]{\includegraphics{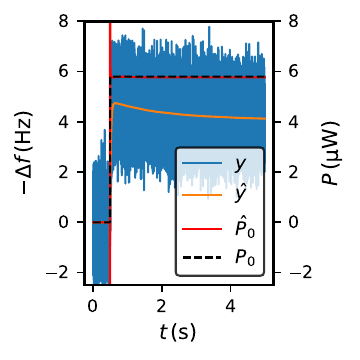}\label{fig:string_kf_step}}

        \subfloat[]{\includegraphics{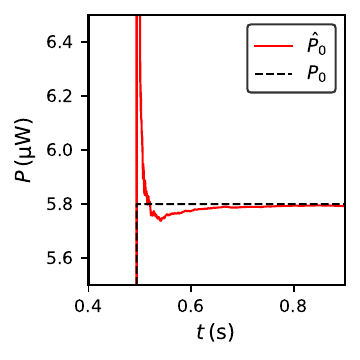}\label{fig:string_kf_step_zoom}} 
        \subfloat[]{\includegraphics{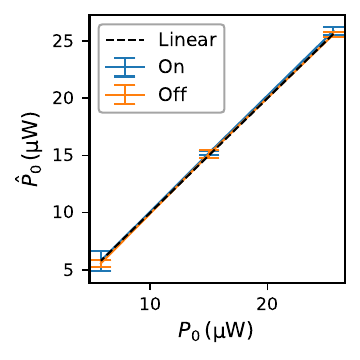}\label{fig:string_kf_stat}}  
        \subfloat[]{\includegraphics{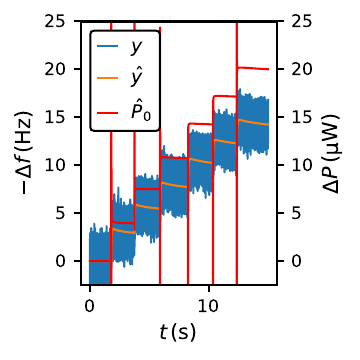}\label{fig:string_kf_stair}}  
        
    \caption{Results of the experiments performed on the string resonator: (a) System identification through calculation and subsequent fitting of the step response to the measured signal. (b) Characterization of measurement noise via the power spectral density of the measured signal, alongside a comparison with the calculated process noise. (Values are displayed in fractional frequency) (c) Visualization of the adaptive Kalman Filter estimates for a measured step response, contrasted against the input laser power to demonstrate stable and precise input power measurement. (d) A detailed view of the transient behavior in the power estimate, highlighting the significant initial estimate error and its rapid convergence to a steady state. (e) A reproducibility and linearity test conducted by executing 5 successive measurements across three power levels, including analyses of both "on" and "off" laser transients. (f) Application of the Adaptive Kalman Filter to a staircase function, showcasing immunity to the system's second slow time constant.
}
    
\end{figure*}

In the first experiment, the performance of the model is evaluated. The thermal model is identified by fitting the parameters to a step response generated by a jump in laser power of $P_0 = 5.8\,\mu\mathrm{W}$. To fit the data, the initial parameters must be determined. In this work, we adopted a three-step approach for data fitting. Initially, we calculated the parameters analytically from the resonator geometry and material properties. Subsequently, a manual fit of the parameters was performed, and in the final step, we ran the \textit{Levenberg-Marquardt} algorithm to obtain the optimal parameters. Table~\ref{tab:param_string} displays the values of the parameters obtained through data fitting, and Figure~\ref{fig:string_fit} shows the step responses for both the analytically calculated and fitted data.
\begin{table*}
    \def\tabW{1.7cm}
    \centering
    \begin{tabular}
    {>{\centering\arraybackslash}m{2cm}  >{\centering\arraybackslash}m{\tabW}  >{\centering\arraybackslash}m{\tabW}  >{\centering\arraybackslash}m{\tabW}  >{\centering\arraybackslash}m{\tabW}  >{\centering\arraybackslash}m{\tabW}  >{\centering\arraybackslash}m{\tabW}  >{\centering\arraybackslash}m{\tabW}  >{\centering\arraybackslash}m{\tabW} }
        \hline
        {\bf Parameter} & $g$ & $C_\mathrm{r}$ & $R_\mathrm{rad}$ & $R_\mathrm{r}$ & $C_\mathrm{f}$ & $R_\mathrm{f}$ & $\alpha_\mathrm{r}$ & $\alpha_\mathrm{f}$ \\
        \hline
        {\bf Calculation} & $357$ & $5.88 \cdot 10^{-10}$ & $3.36 \cdot 10^8$ & $1.48 \cdot 10^8$ & $1.48 \cdot 10^{-7}$ & $1.48 \cdot 10^8$ & $1 \cdot 10^{-6}$ & $1.6 \cdot 10^{-6}$ \\
        \hline
        %{\bf Manual} & $6 \cdot 10^7$ & $2.94 \cdot 10^{-10}$ & $3.36 \cdot 10^8$ & $1.48 \cdot 10^8$ & $4.96 \cdot 10^{-7}$ & $3.72 \cdot 10^8$ & $1 \cdot 10{-6}$ & $2.6 \cdot 10{-6}$ \\
        %\hline
        {\bf Fit} & $357$ & $2.39 \cdot 10^{-10}$ & $3.1 \cdot 10^8$ & $1.45 \cdot 10^8$ & $6.88 \cdot 10^{-7}$ & $2.6 \cdot 10^7$ & $9.89 \cdot 10^{-7}$ & $1.55 \cdot 10^{-6}$ \\
        \hline
    \end{tabular}
    \caption{Comparison between analytically calculated and fitted parameters.}
    \label{tab:param_string}
\end{table*}

The system's measurement noise is characterized by performing a $30\,\mathrm{s}$ measurement at steady state with the laser turned off. The BPF $3\,\mathrm{dB}$ bandwidth and the low-pass filter in the frequency counter are both set to $1\,\mathrm{kHz}$, chosen to be at least one order of magnitude larger than any transient phenomena in the system. The sampling rate of the frequency counter is set to approximately $20\,\mathrm{kHz}$. From the acquired data, a Power Spectral Density (PSD) is generated, as shown in Figure~\ref{fig:string_PSD}. For better visualization, all PSDs are displayed in fractional frequencies. By fitting \eqref{eq:S_v} to the PSD the measurement noise is calibrated to be $S_\mathrm{V} = 5 \times 10^{-16}\,\mathrm{Hz}^{-1}$, while the thermal noise floor is calculated to be $\bar{S}_\mathrm{\Delta T_\mathrm{r}} = 6 \times 10^{-17}\,\mathrm{Hz}^{-1}$.

After the system is characterized by the step response and the noise by the PSD, the adaptive Kalman filter (AKF) can be employed. Figure~\ref{fig:string_kf_step} illustrates the AKF's response to a laser power jump of $P_0 = 5.8\,\mu\mathrm{W}$, induced by controlling the Electro-Optic Modulator (EOM). At the moment of the step, the covariance matrix of the AKF is increased by a factor greater than $10^4$, significantly expanding its bandwidth. This modification renders the system more sensitive to noise, resulting in a highly inaccurate initial estimate. Subsequently, the AKF swiftly decreases the bandwidth, allowing it to rapidly converge to a steady state. In this state, the measurement estimate aligns with the measurement signal trajectory, and the input estimate accurately reflects the step applied to the resonator. Figure~\ref{fig:string_kf_step_zoom} presents a closer view of the input step estimate's transient behavior, demonstrating that approximately $100\,\mathrm{ms}$ to $200\,\mathrm{ms}$ are required for the input estimate to stabilize at the input signal power value.

In the subsequent experiment, the reproducibility of the system is assessed. For three distinct power levels ($5.8\,\mu\mathrm{W}$, $15\,\mu\mathrm{W}$, and $25.6\,\mu\mathrm{W}$), the laser is alternately turned on and off five times, with a $50\,\mathrm{s}$ interval between each sample. An estimate of the input power is collected $200\,\mathrm{ms}$ after the laser event. Figure~\ref{fig:string_kf_stat} displays the results, revealing that the system exhibits almost perfect linear behavior in both "on" and "off" states. The standard deviation among the samples is approximately 1\% of the input power, underscoring the system's high reproducibility. This is particularly noteworthy considering the significant potential for drift and laser power fluctuations due to the lengthy intervals between sample acquisitions.

In the final experiment, the bias of the laser diode is adjusted to incrementally increase its output power in a step-wise manner. Altering the power in such a manner results in degradation of the step edges, attributable to the internal power regulation circuitry of the laser diode, which causes slow edges and edges with overshoots. As observed from Figure~\ref{fig:string_kf_stair}, despite the transients not being perfectly consistent, the power estimation closely resembles a staircase function.

\subsection{NEMS-IR Setup}
\begin{figure*}
    \centering
    \captionsetup[subfigure]{position=top,singlelinecheck=off,justification=raggedright}

        \subfloat[]{\includegraphics{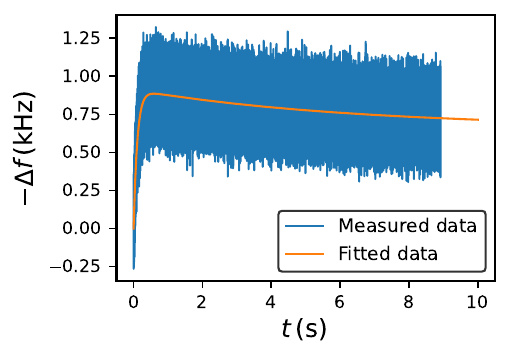}\label{fig:step-fit}}
        \hspace{0.5cm}
        \subfloat[]{\includegraphics{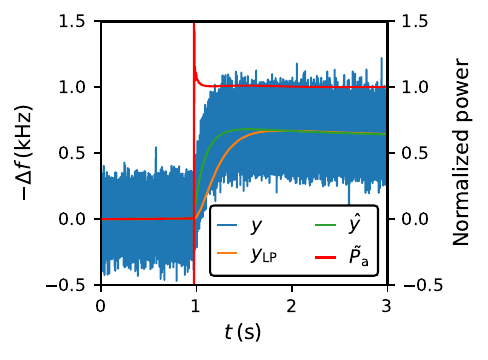}\label{fig:kalman_step}}

        \vspace{0.5cm}
        \subfloat[]{\includegraphics{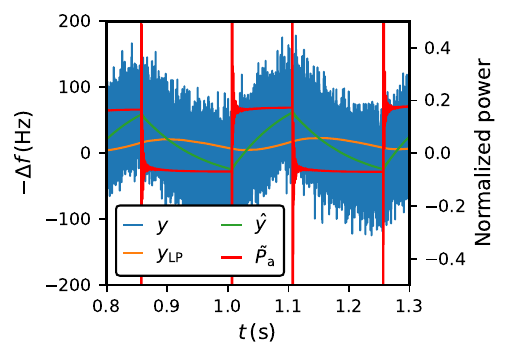}\label{fig:kalman-scan}}
        \hspace{0.5cm}
        \subfloat[]{\includegraphics{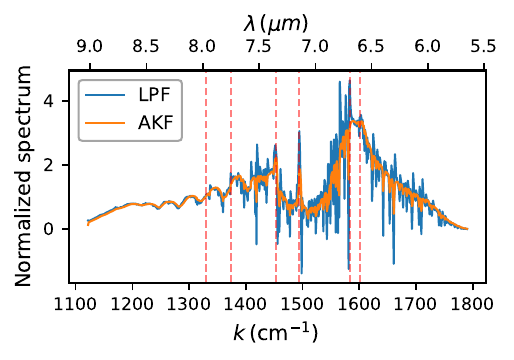}\label{fig:kalman-spectrum}}

    \caption{Results from the NEMS-IR experiments: (a) System identification for the NEMS-IR setup. (b) Adaptive Kalman filtering applied to a laser power step, illustrating a rapid power estimate, a comparatively slower measurement estimate, and an even slower response from a low-pass filter possessing equivalent filtering strength to the AKF. (c) A detailed view of a segment from the QCL step scan, comparing AKF and low-pass filter data. (d) Comparison of the polystyrene spectrum obtained using the adaptive Kalman filter estimate and the low-pass filter, with red lines highlighting the characteristic peaks. (some axes are inverted to provide an easier visualization)
}

\end{figure*}

At the outset of the experiment, it is imperative to characterize the model of the chip. Initially, the QCL is activated. Employing techniques described in Section~\ref{sec: sys_ident}, the transient step response is fitted to the measured data, as depicted in Figure~\ref{fig:step-fit}. Analysis of the data reveals two distinct time constants: a rapid transient in the positive direction and a markedly slower transient in the negative direction. To account for varying absorbance of the structures at different laser wavenumbers, all laser powers in the subsequent experiments will be normalized as
\begin{equation}
    \tilde{P}_a = \frac{\hat{P}_a}{P_{1300}},
\end{equation}
where $P_{1300}$ is the measured power at at a wavenumber of \( k = 1300\,\mathrm{cm^{-1}} \) and $\tilde{P}_a$ the normalized absorbed power estimate.

In the initial experiment, the QCL is set to a wavenumber of $1300\,\mathrm{cm}^{-1}$ with a power output of $0.45\,\mathrm{mW}$. Upon activation, the laser illuminates the NEMS device, and the ensuing data are processed using both the Adaptive Kalman Filter (AKF) and a low-pass filter (LPF), with the cut-off frequency of $1\,\mathrm{Hz}$, possessing a comparable filtering strength (approximately $0.4\,\mathrm{Hz}$ standard deviation over a $300\,\mathrm{ms}$ recording period). As illustrated in Figure~\ref{fig:kalman_step}, the data refined by the AKF exhibit a significantly quicker response compared to that processed by the low-pass filter. Upon detecting a jump, the Kalman filter's bandwidth expands substantially, facilitating a rapid response. Subsequently, the bandwidth narrows swiftly, and the data converge to a smoother curve that represents the optimal estimate. Figure~\ref{fig:kalman_step} also reveals that the estimated absorbed power reacts the most promptly. For models that are both stable and reliable, the power estimate is deemed suitable. Note that any mismatch between the model and the actual experimental system may be mapped into wrong estimates by the AKF trying to reconcile model and measurements.

In the second experiment, the chip is coated with polystyrene, and data collection spans a duration of 3 minutes. The QCL operates in step scan mode, systematically covering the wavenumber range from $1789\,\mathrm{cm^{-1}}$ to $1122\,\mathrm{cm^{-1}}$ in increments of $1\,\mathrm{cm^{-1}}$. For each wavenumber, the laser is activated for $100\,\mathrm{ms}$ and subsequently deactivated for $150\,\mathrm{ms}$.

Figure~\ref{fig:kalman-scan} presents a section of the time-series data obtained from the scan. Notably, the input estimates rapidly converge to a stable, flat line. In contrast, the measurement estimate closely and smoothly tracks the measurement data. Although the low-pass filter succeeds in attenuating noise, it concurrently dampens the signal as well. This observation underscores the AKF's significant enhancement of the system's responsiveness. Utilizing a conventional low-pass filter necessitates a waiting period that is 5 to 10 times longer than the resonator's time constant to allow the signal to stabilize. A further complication arises due to frame stretching, which, influenced by a notably sluggish time constant, alters the signal. This effect complicates the determination of an optimal sampling point for the signal.

To generate the spectrum, data is collected differentially by subtracting consecutive measurements corresponding to the laser's on and off states. This method ensures that the resulting spectrum remains unaffected by slow noise processes, such as thermal drift and random walk processes. Figure~\ref{fig:kalman-spectrum} displays the resultant spectra for the AKF and LPF, which prominently feature the characteristic peaks of polystyrene. It can be observed that the spectrum generated by the AKF has less noise and is cleaner, in comparison to the one generated by the LPF data.

\section{Conclusions} \label{sec:conclusions}
In this study, we aimed to enhance the speed and precision of nanomechanical photothermal sensors by developing a comprehensive thermal heat transfer and noise model suited for optimal Kalman filtering. We validated our model through experiments on a silicon-nitride string resonator and applied it to photothermal IR spectroscopy using a more complex nanoelectromechanical drumhead resonator.

Our thermal model accurately captures system dynamics, including the significant influence of the thermal expansion coefficients' ratio on the frequency response. The adaptable Kalman filter (AKF) we developed provides rapid and precise laser power estimation, suitable for real-time FPGA implementation or offline use. Experimental results demonstrated the model's accuracy and the AKF's effectiveness in managing measurement noise and maintaining stable power estimation.

In IR spectroscopy applications, the AKF enabled fast data sampling and improved precision by eliminating slow time constant issues and successfully identifying characteristic spectral peaks. Overall, our comprehensive thermal model and Kalman-filter-based approach significantly advance the performance of nanomechanical photothermal sensors. The enhanced speed and precision achieved through this work offer promising avenues for further research and practical applications in the field of nanomechanical sensing. Future efforts will focus on refining the model and expanding its applicability to other resonator types and sensing modalities.

\section*{Acknowledgements} \label{sec:acknowledgements}
This work received funding from the European Innovation Council under the European Union's Horizon Europe Transition Open program (Grant agreement: 101058711-NEMILIES).
This project received funding from Defense Advanced Research Projects Agency (DARPA) Optomechanical Thermal Imaging (OpTIm) Technical Area (TA) 1 Broad Agency Announcement (BAA), HR001122S0055.
The project received funding from the Novo Nordisk Foundation for the project MASMONADE with the grant reference number NNF22OC0077964.

\bibliography{bibliography.bib}

\end{document}